\input psfig.sty

\documentclass{aa}

\begin{document}

\thesaurus{}

\title{The Vela pulsar proper motion revisited with HST 
astrometry\thanks{Based 
on observations with the NASA/ESA Hubble Space Telescope,
obtained at the Space Telescope Science Institute, which is operated by AURA,
Inc.  under contract No NAS 5-26555}.}

\author{A. De Luca\inst{1,2} \and R.P. Mignani\inst{3} \and P.A. 
Caraveo\inst{2,4} 
}

\institute{Universit\`a di Pavia, Dipartimento di Fisica Nucleare e Teorica, 
Via Bassi 6, I-27100 Pavia (Italy)  \and Istituto di Fisica Cosmica del CNR ``G.  Occhialini", Via Bassini
15, I-20133 Milan, Italy  \and STECF-ESO, Karl Schwarzchild Str. 2, D-85740
Garching bei Muenchen, Germany  \and Istituto Astronomico, Via Lancisi 29,
I-00161 Rome, Italy }

\date{Received / Accepted } \offprints{P.A.Caraveo (pat@ifctr.mi.cnr.it)}

\titlerunning{Vela proper motion revisited with HST}
\maketitle

\begin{abstract}

Using all the HST  WFC/WFPC2 images of the field  collected so  far, we
have performed an accurate  relative astrometry analysis to  re-assess
the value of the  Vela pulsar proper  motion. Although covering a much
shorter time span, our measurement clearly
confirms the previous result  obtained by Nasuti  et al.  (1997) using
ground-based optical  data and nails the  proper motion value to $\mu=
52\pm3 $ mas $yr^{-1}$. 


\end{abstract}

\section{Introduction}

The Vela pulsar is one  of the isolated  neutron stars with the widest
observational  database,  spanning from  radio   waves  to high-energy
$\gamma$-rays.  Nevertheless, the value  of  its distance is still  in
debate.  The  canonical value of 500  parsec,  derived by Milne (1968)
from  the  radio signals    dispersion  measure,  has  been   recently
questioned by several independent investigations.  Studies of both the
kinematics of the host supernova remnant (Cha et al. 1999; Bocchino et
al. 1999)  and constraints on the  neutron star radius imposed  by the
pulsar    soft X-ray spectrum  (Page  et  al.   1996) have suggested a
significant downward revision of the distance.  In both cases, a value
of  $\approx$ 250   parsec appears  more   likely than the canonical one. 
 A   model-free
evaluation of  the  distance which  could settle  the question  can be
obtained only by measuring the annual parallax  of the pulsar. To this
aim we have been granted a  triplet  of consecutive HST/WFPC2 
observation  to  be  performed six
months apart at the epochs of the maximum parallactic displacement.

\begin{table} \begin{tabular}{c|c|c|c} \hline Optical & $\mu$ &
$\mu_{\alpha}\cos{\delta}$ & $\mu_{\delta}$ \\ & mas~yr$^{-1}$ & mas~yr$^{-1}$ 
& mas~yr$^{-1}$ \\ \hline
\hline Bignami \& Caraveo & $<60$ & - & - \\ 1988 & & & \\ \hline \"Ogelman et
al.  1989 & $38\pm8$ & $-26\pm6$ & $28\pm6$ \\ \hline Markwardt \& \"Ogelman &
$49\pm4$ & $-41\pm3$ & $26\pm3$ \\ 1994 & & & \\ \hline Nasuti et al.  1997 &
$52\pm5$ & $-47\pm3$ & $22\pm3$ \\ \hline \hline Radio & & & \\ \hline Bailes 
et
al.  1989 & $49\pm4$ & $-40\pm4$ & $28\pm2$ \\ \hline Fomalont et al.  1992 &
$116\pm62$ & $-67\pm20$ & $-95\pm75$ \\ \hline Fomalont et al.  1997 & 
$53\pm19$
& $-50\pm5$ & $-16\pm18$ \\ \hline \end{tabular} \caption{Previous estimates of
the Vela pulsar proper motion in literature} \end{table}

In this  paper we present a first  result of our program:  an accurate
re-assessment of the pulsar proper motion.  This is another open point
in the Vela   pulsar phenomenology:  although certainly  present,  its
actual value is still uncertain.  Previous estimates obtained through radio and
optical observations   led to  conflicting  results (see  Tab.1 for  a
summary), spanning from 38 mas~yr$^{-1}$  (\"Ogelman et al.1989) to 116
mas~yr$^{-1}$ (Fomalont et al.   1992).   These discrepancies are  due
both to the rather poor angular resolution of the first optical images
of the field,  which reduced the  accuracy of relative astrometry, and
to the timing irregularities of the Vela pulsar signal, which possibly
affected the   reliability  of radio   positions. However,  the  newly
operational southern  VLBI has already   been used on the Vela  pulsar
yielding preliminary results of vastly improved accuracy (Legge, 1999).

\begin{table*} \begin{tabular}{c|c|c|c|c|c|c} \hline & Date &
Instrument & Pixel size (arcsec) & Filter & N$^{\circ}$ exposures & Exposure 
(s) \\
\hline\hline 1 & 1993 February 26 & WFC & 0.1 & F555W & 4 & 400 \\ \hline 2 & 
1997 June
30 & WFPC2 & 0.045 & F555W & 2 & 1300 \\ \hline 3 & 1998 January 2 & WFPC2 & 
0.045 &
F555W & 2 & 1000 \\ \hline 4 & 1999 June 30 & WFPC2 & 0.045 & F555W & 2 & 1000 
\\
\hline \end{tabular} \caption{HST observations of the Vela pulsar
field. The columns give the image number, the observing epoch, the instrument 
used, the
pixel size in arcsec, the filter, the total number of repeated
exposures for each observation and the exposure duration in seconds.}
\end{table*}

\section{The data analysis}

The best way to gauge the angular displacement of an object between different
epochs is to perform relative optical astrometry measurements. 
This needs: 

\begin{itemize}
\item a set of ``good'' reference stars, accurately positioned,
 to provide the relative reference frame  for each image 
\item a reliable procedure to align the reference frames 
\end{itemize}

In our relative astrometry analysis we have used all the images of the
Vela pulsar field collected so far by the HST (see Tab.2).
The observations, all taken through the F555W ``V filter'', have been
obtained either using the original WFPC and the WFPC2, with the pulsar
positioned in one of the WFC chips or in the PC one, respectively. 
Observation \#1 has been retrieved from the ST-ECF database and
recalibrated on-the-fly by the archive pipeline using the most recent
reference data and tables, while observations \#2 and \#3 were obtained as 
part of our original parallax program, included in cycle 6 but never completed.
The program is now being repeated in cycle 8 and image \#4 is 
indeed the first of the new triplet of  WFPC2 observations.
For each epoch, cosmic-ray free
images were obtained by combining coaligned exposures.

\begin{figure}
\centerline{\hbox{\psfig{figure=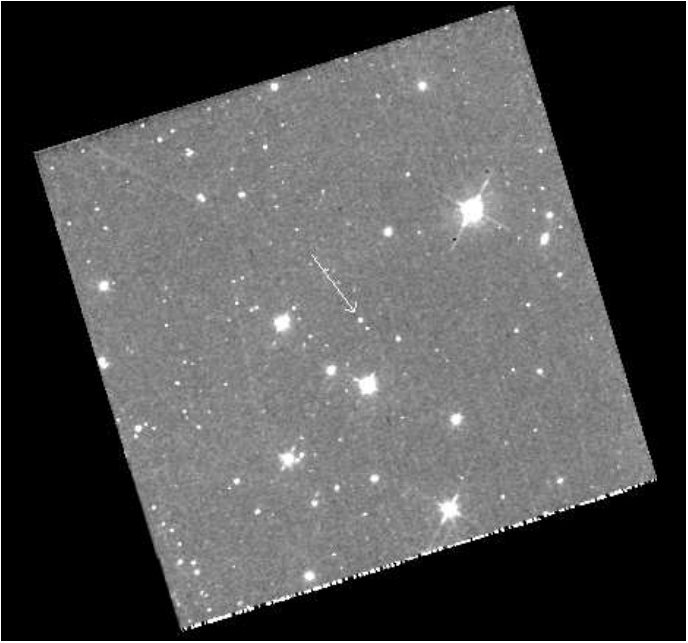,height=85mm,clip=}}}
\caption{Planetary Camera image of the Vela pulsar field taken in 1997
through the F555W filter. North is up, East is left. 
The arrow marks the position of the Vela pulsar.}
\end{figure}

The choice of  reference stars was limited to  the chip containing the
pulsar optical counterpart   (see Caraveo   \&   Mignani 1999 for    a
qualitative discussion of the inter-chip astrometry).  25 common stars
were selected in the PC images of 1997, 1998 and 1999; 19 of them were
identified in the  1993 WFC image.  This  defines our set of reference
stars.  Their coordinates were calculated
by 2-D gaussian  fitting of the  intensity profile.  The evaluation of
the positioning  errors was conservative; for  each reference star the
fit was repeated several times on  centering regions of growing areas,
till  errors showed no  more  dependence on the background  conditions
(see Caraveo et  al. 1996).  Uncertainties  on the  centroid positions
were typically of order 0.02$\div$0.06 pixel  (i.e.  1$\div$3 mas) per
coordinate  in   the  WFPC2 images   and 0.03$\div$0.06   pixel  (i.e.
3$\div$6 mas) per coordinate in the WFC one. \\
The  values of   the coordinates of    reference stars have been  then
corrected for the effects of the significant geometrical distortion of
the WFC and the  WFPC2 CCDs. This correction  has been applied following two
different mappings of  the instrument field of  view: (i) the solution
determined by Gilmozzi et al.  (1995), implemented  in the STSDAS task
{\em  metric} and  (ii) the solution   determined  by Holtzman et  al.
(1995). The procedures turned out to be equivalent, with the rms
residuals on the reference stars coordinates after image superposition
(see below) consistent within few tenths of mas. \\
As a reference for the image superposition, we have choosen the 1997 one. 
Given the abundance of reference stars, the alignement has been
performed following  the traditional astrometric  approach, consisting
of a linear transformation with 5 free parameters i.e. 2  independent
translation  factors, 2 scale  factors for $X$ and  $Y$ and a rotation
angle.  The residuals of reference stars positions clustered around 0
and appeared randomly distributed on the field  of the image, showing no
systematic  effect. Thus, we are confident   that our analysis is bias
free and   reliable.  As  expected, the  average  rms  residual on the
reference stars positions is higher in the WFC-to-PC superposition (of
order 4 mas) than in the PC-to-PC case (of order 2 mas). 
We  note  that in all  cases  the  relative orientation of  the images
coincide, within few hundredths of degree, with the difference between
the corresponding telescope roll angles. 
The alignement procedure has been repeated after applying different
image sharpening algorithms (e.g. gaussian filtering), 
yielding very similar results. 

 \begin{figure*}[t]
\centerline{\hbox{\psfig{figure=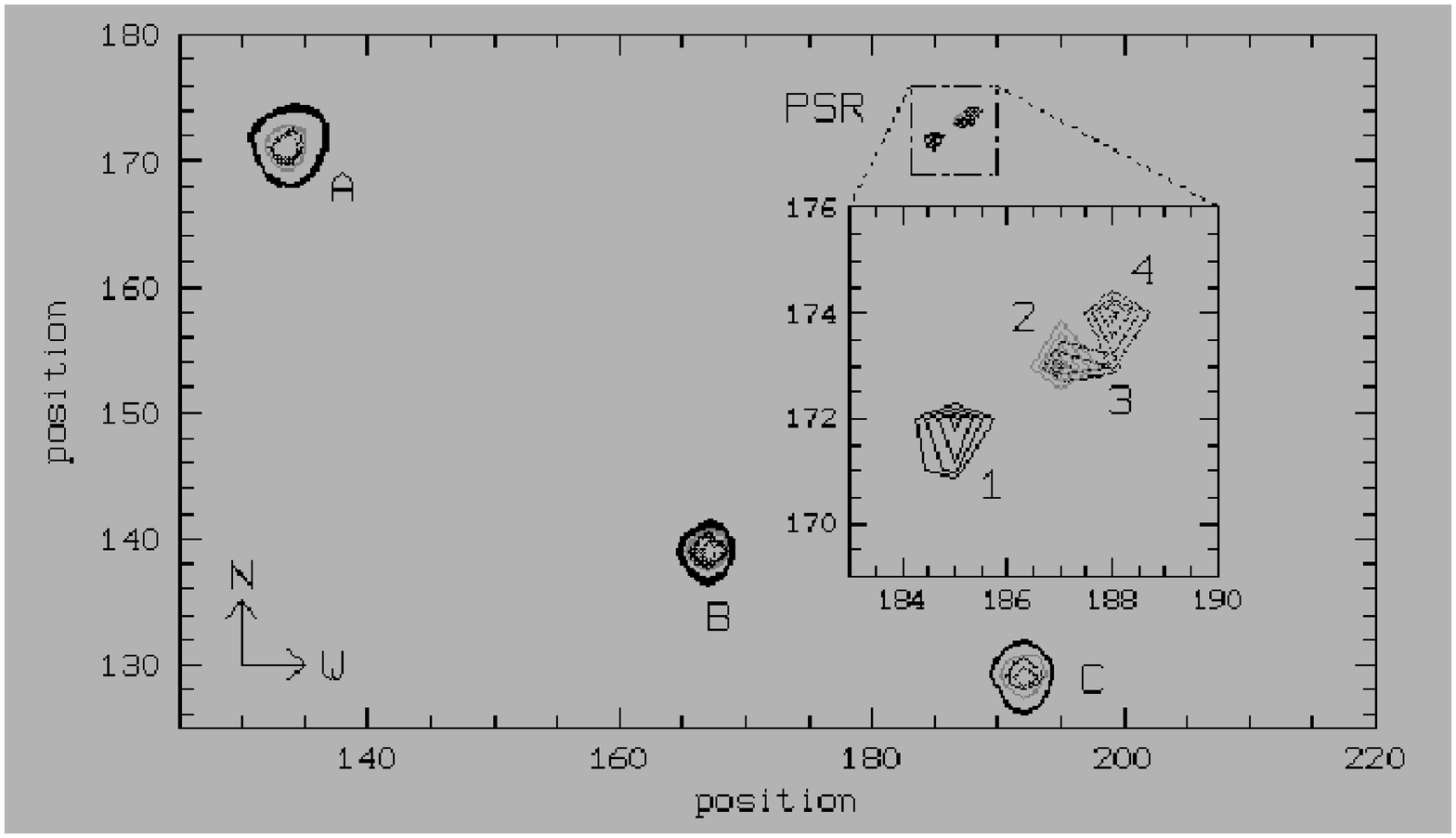,height=8cm,clip=}}}
\caption{Contour plot obtained from the four images of Tab.1, after 
the superposition procedure described in the text. 
For clarity, the images have been finally aligned in RA and Dec
according to the roll angle of image 2. Axis units are 
PC pixels (0\farcs045). The objects labelled as A, B,
C, easily identifiable in Fig.1, are three of the reference stars used
to compute the image superposition. The pulsar environment has been 
enlarged for clarity in the inset, where numbers refer to Tab.2. 
}
\end{figure*}

\section{The proper motion}

The displacement of the Vela pulsar position
can be immediately appreciated in the isophotal plot of Fig.2, which shows 
the zoomed superposition of all the frames taken at the different epochs. 
To measure  the pulsar proper motion, we have  performed
linear fits to its RA/Dec displacements vs time.  As  a first step, we
used all the available  points to  have a  longer time span.
However, owing to  the coarser angular  resolution of the  1993 image,
taken with the  WFC,  almost  indistinguishable
results are obtained using only  the
1997-1999  points,  all obtained  with  the PC. We also tried a direct 
comparison of the 1997
and 1999 points, taken  exactly at the same  day of the year i.e. with
identical parallax  factors,  which should lead to the clearest measurement of  the
proper  motion.  All of  the steps described above  yielded 
results  largely consistent   within  the errors.  
%However,  it  is
%worth noting that the rejection  of the 1998  image yields a reduction
%of the ${\chi}^2$ value in the 1997-1999  Dec fit.  In particular, the
%1997-1998 Dec  displacement is  below   the trend  resulting from  the
%global analysis, hinting the presence of a measurable annual parallax. 

Since all the proper motion values obtained through different image
processing/frame superposition/displacement fit are consistent 
within $\approx 2$ mas~yr$^{-1}$ from each other, 
we have conservatively assumed the value of 2 mas ~yr$^{-1}$ (per
coordinate) as our overall error estimate.
Thus, our final estimate of the Vela pulsar proper motion is:

$$ \mu_{\alpha}\cos{\delta}=\hspace{2mm} -46\pm 2 \hspace{2mm}
mas~yr^{-1}$$ 

$$
\mu_{\delta}=\hspace{2mm}24\pm 2 \hspace{2mm} mas~yr^{-1}$$ for a total proper 
motion
of:  $$ \mu=\hspace{2mm}52\pm 3 \hspace{2mm} mas~yr^{-1}$$ with a position 
angle of
$297^{\circ} \pm 2^{\circ}$.

\section{Conclusions}

Analysis of  HST data yielded  the most  accurate measure of  the Vela
pulsar proper motion. Our value is in excellent agreement with the one
of Nasuti  et al.  (1997),but the  errors are now reduced.  What is
worth mentioning here  is that our   data refer to  a time  span much
shorter than the 20  years interval of  the previous work. Indeed, two
years of HST observations are vastly enough to improve over 20
years observations from the ground. Once more,
this is a clear demonstration  of the excellent potentialities of  HST
astrometry. \\ Transforming  proper  motion into  a transverse  velocity
requires the knowledge of   the object's distance. At the   canonical,
although uncertain,  500  pc distance  the implied velocity   would be
$\simeq 130$~km~s$^{-1}$, somewhat on the low side for the fast moving
pulsar  family (Lorimer, Bailes \& Harrison  1997).  A reduction of
the pulsar distance will similarly reduce the transverse speed to an
embarassingly low value.  Nailing down the
pulsar  proper  motion  is  the  first  step  to   assess  the  annual
parallactic displacements of the  source.  Our next observations will,
hopefully, yield a direct measurement of the Vela pulsar distance. 

\begin{acknowledgements}
Part of this work was done at the ST-ECF of Garching. ADL wishes to thank
ECF for the hospitality and the support during that period.
ADL is sincerely grateful to the Collegio Ghislieri of Pavia (Italy), to
the Pii Quinti Sodales association of Pavia and to the Maximilianeum Stiftung
of Munich, which offered a truly essential support and a warm hospitality
for his stay in Germany.
\end{acknowledgements}

\end{document}